\documentclass[aps,prl,twocolumn,groupedaddress,showpacs]{revtex4-1}
\pdfoutput=1
\usepackage{graphicx}
\usepackage{amssymb}
\usepackage{bm}

\usepackage{pdfpages}
\usepackage{pgffor}

\newcommand{\corr}{ \mathcal{G}}
\newcommand{\Del}{\delta}

\newcommand{\alphaIn}{\mathcal{E}}

\newcommand{\vC}{\tilde{v}}
\newcommand{\normord}[1]{:\mathrel{#1}:}
\newcommand{\suppref}[1]{} %in order to not make mistake in supplement Eq's numeration 
\newcommand{\ket}[1]{\ensuremath{\left|#1\right\rangle}}  							 	% |...>
\newcommand{\integralb}[3]{\int\limits_{#1}^{#2} \! \mathrm{d} #3\,} 	% integral
\newcommand{\integral}[1]{\int \! \mathrm{d} #1\,}                    % integral without limits
\newcommand{\rs}{\rm \scriptscriptstyle}
\def\nn{\nonumber}
\newcommand{\ra}{\ensuremath{\rightarrow}\xspace}
\newcommand{\dr}{^{\dagger} }

\def\a{\ensuremath{\alpha}\xspace}
\def\Schroedinger{Schr\"{o}dinger\xspace}
\def\beqa{\begin{eqnarray}}
\def\eeqa{\end{eqnarray}}
\newcommand{\meanv}[1]{\left\langle#1\right\rangle}							% <...>

\def\aOp{{\hat{a}_u}}%{a_u}
\def\psiTwoB{\phi} %{\Upsilon}
\def\phiTwo{\varphi}
\def\phiTwoMass{\vartheta}

\newcommand{\beginsupplement}{%
        \setcounter{table}{0}
        \renewcommand{\thetable}{S\arabic{table}}%
        \setcounter{figure}{0}
        \renewcommand{\thefigure}{S\arabic{figure}}%
     }

\usepackage{xspace} % add trailing space after commands like \gpe
\usepackage{color}
\usepackage{verbatim}

\begin{document}
%\draft
%\preprint{{\bf ETH-TH/98-??}}

\title {Quantum theory of Kerr nonlinearity with Rydberg slow light polaritons}

\author{Przemyslaw~Bienias}
\affiliation{Institute for Theoretical Physics III and Center for Integrated Quantum Science and Technology, University of Stuttgart, Pfaffenwaldring 57, 70550 Stuttgart, Germany}
\author{Hans~Peter~B\"{u}chler}
\affiliation{Institute for Theoretical Physics III and Center for Integrated Quantum Science and Technology, University of Stuttgart, Pfaffenwaldring 57, 70550 Stuttgart, Germany}

\date{\today}

\begin{abstract}
We study the propagation of Rydberg slow light polaritons through an atomic medium for intermediate interactions.  Then, the dispersion relation for the 
polaritons is well described by the slow light velocity alone, which allows for an analytical solution for arbitrary shape of the atomic cloud. We demonstrate the connection of  Rydberg polaritons to the behavior of a conventional  Kerr nonlinearity for weak interactions, and determine the leading quantum corrections  for increasing interactions. We propose an experimental setup which allows one to measure the effective two-body interaction potential  between slow light polaritons as well as higher body interactions.
\end{abstract}

%42.65.Hw Phase conjugation; photorefractive and Kerr effects
%32.80.Ee Rydberg states
%42.50.Gy Effects of atomic coherence on propagation, absorption, and amplification of light; electromagnetically induced transparency and absorption
%42.50.Nn Quantum optical phenomena in absorbing, amplifying, dispersive and conducting media; cooperative phenomena in quantum optical systems

\maketitle

Photons interact with its environment much weaker than other quanta and therefore represent an excellent carrier of information. 
On the other hand, a long-standing goal is the realization of a strong and controllable interactions on the level of  individual photons.
Such an interaction  would pave the way towards ultralow-power all-optical signal processing \cite{Hu2008a,Miller2010}, which in turn has important applications
in  quantum information processing and communication \cite{Milburn1989,Kimble2008,Muthukrishnan2004,Giovannetti2011}. 
 A natural mechanism  for an interaction is provided by  the Kerr nonlinearity of conventional materials \cite{Boyd2003a}, but unfortunately 
 is restricted to high intensities of the fields \cite{Matsuda2009}. On the other hand,
 the appearance of a strong interaction between individual photons has been experimentally realized using  Rydberg slow 
 light polaritons. Here, we provide the theoretical framework to connect this regime of 
 strong interaction with the  phenomena of a classical Kerr nonlinearity.

Rydberg slow light polaritons have emerged as a highly promising candidate to engineer strong interactions between optical photons
with a tremendous recent experimental success. %~\cite{Peyronel2012a,Maxwell2013,Firstenberg2013a,Gorniaczyk2014,Tiarks2014,Gorniaczyk2015,Tresp2015}. 
A variety of applications were shown such as a deterministic single 
photon source \cite{Dudin2012}, an atom-photon entanglement generation \cite{Li2013b}, as well as a single photon switch \cite{Baur2014} and  transistors \cite{Gorniaczyk2014,Tiarks2014,Gorniaczyk2015}. Moreover,  the regime of strong interaction between %copropagating 
photons has been experimentally demonstrated leading 
to a medium transparent only to single photons \cite{Peyronel2012a}, as well as the appearance of bound states for photons \cite{Firstenberg2013a}. From theoretical point of view, 
the effective low energy theory is well  understood from a microscopic approach \cite{Otterbach2013,Bienias2014b}, 
but a full description of the propagation of photons through the medium is limited to extensive numerical simulations and low photon number~\cite{Gorshkov2011a,Peyronel2012a,Firstenberg2013a,He2014a,Moos2015a,Maghrebi2015,Maghrebi2015c}.

\begin{figure}%[H]
\includegraphics[width= 0.95\columnwidth]{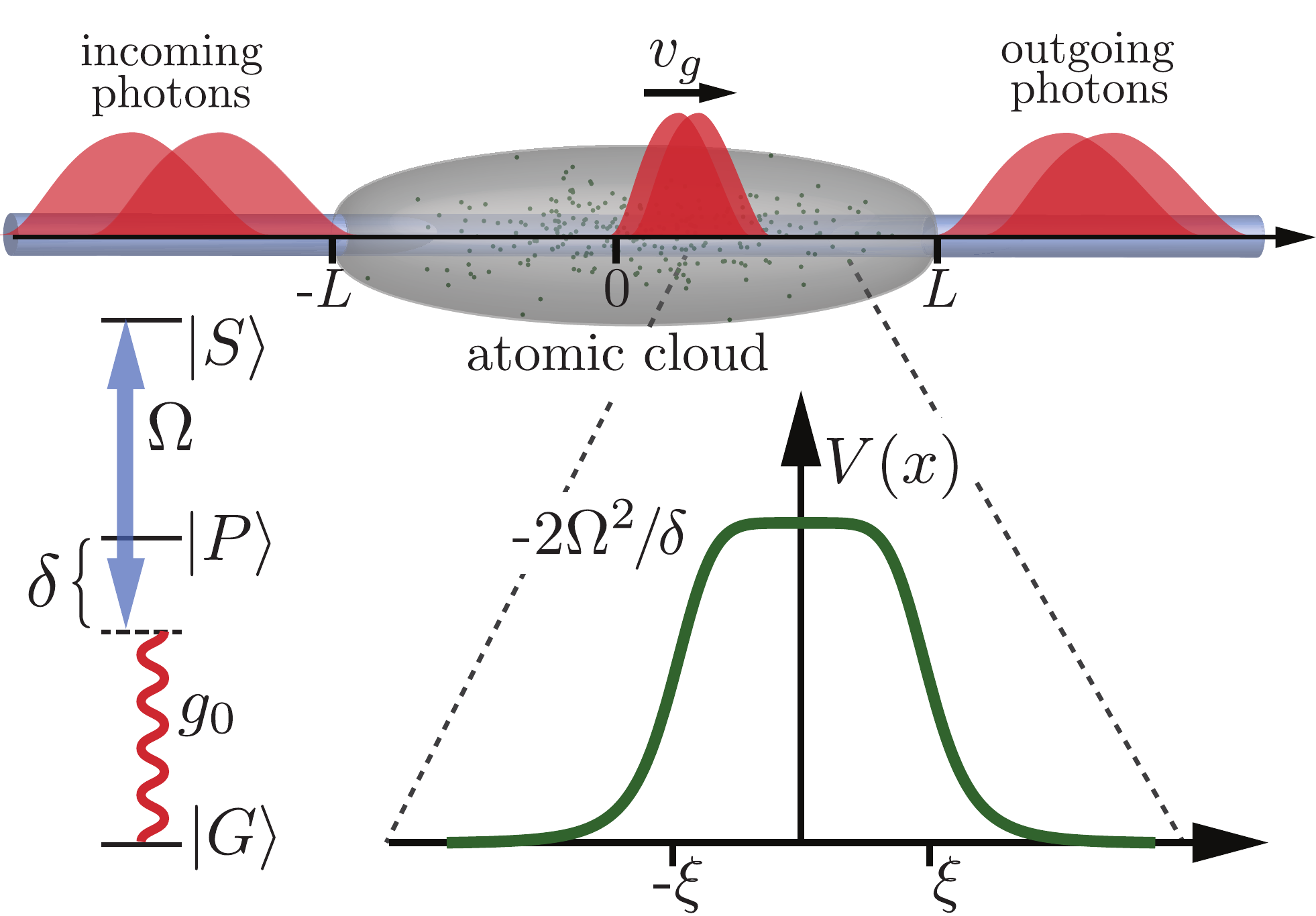}
%\includegraphics[width= 0.95\columnwidth]{beam_through_cloud_9b_v4_broader_blue}
%fig_propagation2_v6a
%}
\caption{
Setup of Rydberg slow light polaritons: each atom consists of three relevant levels, ground state $\ket{G}$, intermediate p-level $\ket{P}$ and Rydberg state $\ket{S}$; the latter are coupled by a strong laser. Incoming photons with a single transverse channel enter the medium and are converted into slow light  $v_g<c$ Rydberg polaritons. The interaction between the Rydberg states provides an effective interaction $V(x)$ for the polaritons.
}
\label{fig1}
\end{figure}

In this letter, we provide the full input-output formalism of Rydberg polaritons for {intermediate} interaction 
strength but arbitrary incoming photon number and shape of the atomic medium. The analysis is performed in the regime with large detuning from 
the intermediate p-level, where losses are strongly suppressed and the effective low-energy theory for the polaritons is well described 
by an effective interaction potential  \cite{Bienias2014b}.  We demonstrate the connection of  Rydberg polaritons to the behavior of a conventional 
Kerr nonlinearity for weak interactions, and determine the leading quantum corrections for such a Kerr nonlinearity. 
We demonstrate the potential to experimentally determine the effective interaction potential as well as higher body interactions between the
slow light polaritons within a  homodyne setup.

It is important to point out that previous approaches to describe the quantum propagation of photons in a nonlinear Kerr medium
based on a quantization of the phenomenological nonlinear equations provide an inconsistent quantum field theory \cite{Blow1991,Haus1992,Joneckis1993}. This inconsistency 
was removed by requiring a non-local response in time, which in addition provides a noise term \cite{Boivin1994,Shapiro2006}. For Rydberg slow light polaritons such 
a non-local response in time is absent, but the microscopic analysis naturally provides a mass term accounting for deviation from the slow 
light velocity, as well as a finite range of the effective interaction potential describing the blockade phenomena.  Both terms alone are sufficient 
to render the quantum theory well defined. As a consequence, we conclude that the proposed inability to generate a photonic phase gate by a large Kerr nonlinearity  \cite{Shapiro2006} 
does not apply to Rydberg slow light polaritons.

We consider a system of Rydberg slow light polaritons in the dispersive limit with large detuning $\Del > \gamma, \Omega$ 
from the intermediate p-level, see Fig.~\ref{fig1}. Here, $\gamma$ describes the decay rate of the p-level, while $\Omega$ denotes
the Rabi frequency of the  coupling laser. Within this regime, losses are strongly suppressed and the intermediate p-level can 
be adiabatically eliminated \cite{Bienias2014b}.  
%Finally, we are interested in the regime of slow light with the collective coupling 
%$g = g_{0} \sqrt{\bar{n}_{a}} \gg \Omega$; here, $g_{0}$ is the single atom coupling strength, 
%while $\bar{n}_{a}$ is the atomic density in the trap center. 
We are interested in the propagation of photons along a one-dimensional mode through the medium with frequency close to the condition of
electromagnetic induced transparency. 
In the regime with a low density  of Rydberg polaritons, the  system is well described by an effective low energy quantum theory \cite{Bienias2014b}.
%In the dilute regime, where the density of Rydberg polaritons is low compared to the range of the interaction potential,
%the  system is well described by an effective low energy quantum theory \cite{Bienias2014b}.
The interaction potential between the polaritons is characterized  by a blockade radius $\xi$  and the potential depth  $2\hbar \Omega^2/\Del$ 
at short distances. For a microscopic van der Waals interaction with $C_{6} \Del < 0$
the effective interaction potential reduces to $ V(x)= -(2 \hbar \Omega^2 /\Del) [1+ (x/\xi)^6]^{-1}$
with the blockade radius $\xi = (|C_{6} \Del|/2\Omega^2)^{1/6}$, see Fig.~\ref{fig1}. 
Note, that for increasing polariton densities  additional many-body interactions are expected to appear \cite{Jachymski2016}. In the following, we mainly focus on
 the two body interactions,  but the extension to include many-body interactions is straightforward and its influence is discussed in the last part. 

%Note that for high polaritonic densities additional many-body interactions are expected to appear. Shapes of these interaction potentials are up to now unknown.
%Therefore, in the first part of the manuscript we focus only on the two body interactions, and in the last part we apply our analysis to many-body interactions.

The kinetic energy for the polaritons at low energies is determined by the slow light velocity of the polaritons and
an effective mass term accounting for the curvature in the dispersion relation. The important aspect for the present
analysis is the possibility to drop the mass term for moderate interactions between polaritons.
The precise condition for the validity of this approximation is  discussed below. Then, the Hamiltonian describing the
propagation  of photons through the spatially inhomogeneous medium with atomic density $n_{a}(x)$  is given by %\cite{Bienias2014b}
%
%Furthermore, for moderate interactions between the polaritons, it is sufficient to consider only the leading order term in 
%the kinetic energy, which reduces to the slow light velocity; especially, we can drop the mass term of the polaritons.
%The precise condition for the validity of this approximation is discussed below. The Hamiltonian describing the propagation 
%of photons through the spatially inhomogeneous medium with atomic density $n_{a}(x)$  is then given by \cite{Bienias2014b}
%
\begin{eqnarray}
   H & = & \integral{x}  \: \left[ \beta(x) \psi^{\dag}(x) \right] \left( - i \hbar c \partial_x \right) \left[ \beta(x)\psi(x)\right]\\
   & +&   \frac{1}{2} \integral{x\, \text{d}y} \, n(x) n(y) V(x-y) \psi^{\dag}(x) \psi^{\dag}(y) \psi(y)\psi(x) \nonumber.
\end{eqnarray}
Here, $\psi$ and $\psi^{\dag}$ denote the bosonic field operators for the Rydberg slow light polaritons 
with $[ \psi(x), \psi^{\dag}(x')] = \delta(x-x')$.  Furthermore, $\beta(x)$ describes the amplitude of the polariton to
be in a photonic state and is related to the slow light velocity $v_{g} = c \beta(x)^2$, while $n(x) = 1- \beta(x)^2$ is
the probability for the polariton to be in the Rydberg state. These quantities are determined by 
the atomic density $n_{a}(x)$ via $\beta(x) = {\Omega}/{\sqrt{\Omega^2 + g_{0}^2 n_{a}(x) }}$ 
with $g_{0}$ the single atom coupling. 
%Here, we allow for a spatially inhomogeneous atomic density $n_{a}(x)$, which allows us to account for the precise spatial atomic distribution.
Note that outside the atomic medium the operator $\psi$ describes non-interacting photons.

In the following, it is convenient to introduce a coordinate transformation which 
removes the reduced velocity $v_{g}$ of the polaritons inside the media, i.e., we measure 
distances in the time $z/c$ which is required for the polaritons to reach the position $x$. 
The coordinate transformation takes the form 
$ z=\zeta^{-1}(x) =\int_{0}^{x}\text {d}y \left({1}/{ \beta(y)^{2}}\right)$, and the Hamiltonian reduces to
\begin{eqnarray}
   H &=& - i \hbar c \integral{z}  \:  \hat{\psi}^{\dag}(z) \partial_z  \hat{\psi}(z)  \label{qmbt}\\
& &+ \frac{1}{2} \integral{z\!\text{ d}w} \tilde{n}(z) \tilde{n}(w) \tilde{V}(z,w) \hat{\psi}^{\dag}(z) \hat{\psi}^{\dag}(w) \hat{\psi}(w)\hat{\psi}(z) \nn
\end{eqnarray}
with $\tilde{n}(z) = n(\zeta(z))$, $\tilde{V}(z,w) = V(\zeta(z)- \zeta{(w)})$, and $\hat{\psi}^{\dag}(z) = {\psi^{\dag}(\zeta(z))}{\beta(\zeta(z))}$; the new operators $\hat{\psi}$ still satisfy the bosonic canonical commutation relations.

The quantum many-body theory in  Eq.~(\ref{qmbt}) is exactly solvable.  This remarkable property is
most conveniently observed by analyzing the Heisenberg equations for the field operator $\hat{\psi}(z,t)$,
%It is a remarkable property of the quantum many-body theory in Eq.~(\ref{qmbt}), that it is exactly solvable. 
%We start with studying the  Heisenberg  equation for the field operator $\hat{\psi}(z,t)$,
%which takes the form
%
\begin{equation}
  i \hbar  \partial_{t}  \hat{\psi}(z,t) =  - i \hbar c \partial_z  \hat{\psi}(z,t)  + K(z,t)   \hat{\psi}(z,t)  
\end{equation}
with the operator $K(z,t)$ accounting for the interaction,
\begin{equation}
  K(z,t) = \integral{w}  \tilde{n}(z) \tilde{n}(w) \tilde{V}(z,w) \hat{\psi}^{\dag}(w,t)  \hat{\psi}(w,t)  .
\end{equation}
In the following, we denote by $\hat{\psi}_{0}(z)$ the 
non-interacting field operator at time $t=0$. Then, the interacting field operator $\hat{\psi}(z,t)$,
satisfying the Heisenberg equation above, reduces to
$ %\begin{equation}
\hat{\psi}(z,t) = e^{-i  \hat{J}(z,t)  } \hat{\psi}_{0}(z-c t)
$ %\end{equation}
with the operator
\begin{equation}
\hat{J}(z,t) = \frac{1}{c \hbar} \int_{z-ct}^{z} \hspace{-15pt}\text{d}w \int_{-\infty}^{\infty}  \hspace{-12pt} \text{d}u \: \tilde{n}(w) \tilde{n}(u) \tilde{V}(u,w) \hat{ I}(z\! -\! w\!+\!u \!-\! c t) 
\label{Jzt}
\end{equation}
and the polariton density operator $\hat{I}(z) = \hat{\psi}^{\dag}_{0}(z) \hat{\psi}_{0}(z)$. 
%The solution requires
%the operator $\exp(-i\hat{J}(z,t))$ to be unitary, and therefore, and requires that
%	the interaction is purely dispersive with strongly suppressed losses.

%\emph{ Two-photon solution:}  

\begin{figure}%[H]
\includegraphics[width= 0.47\columnwidth]{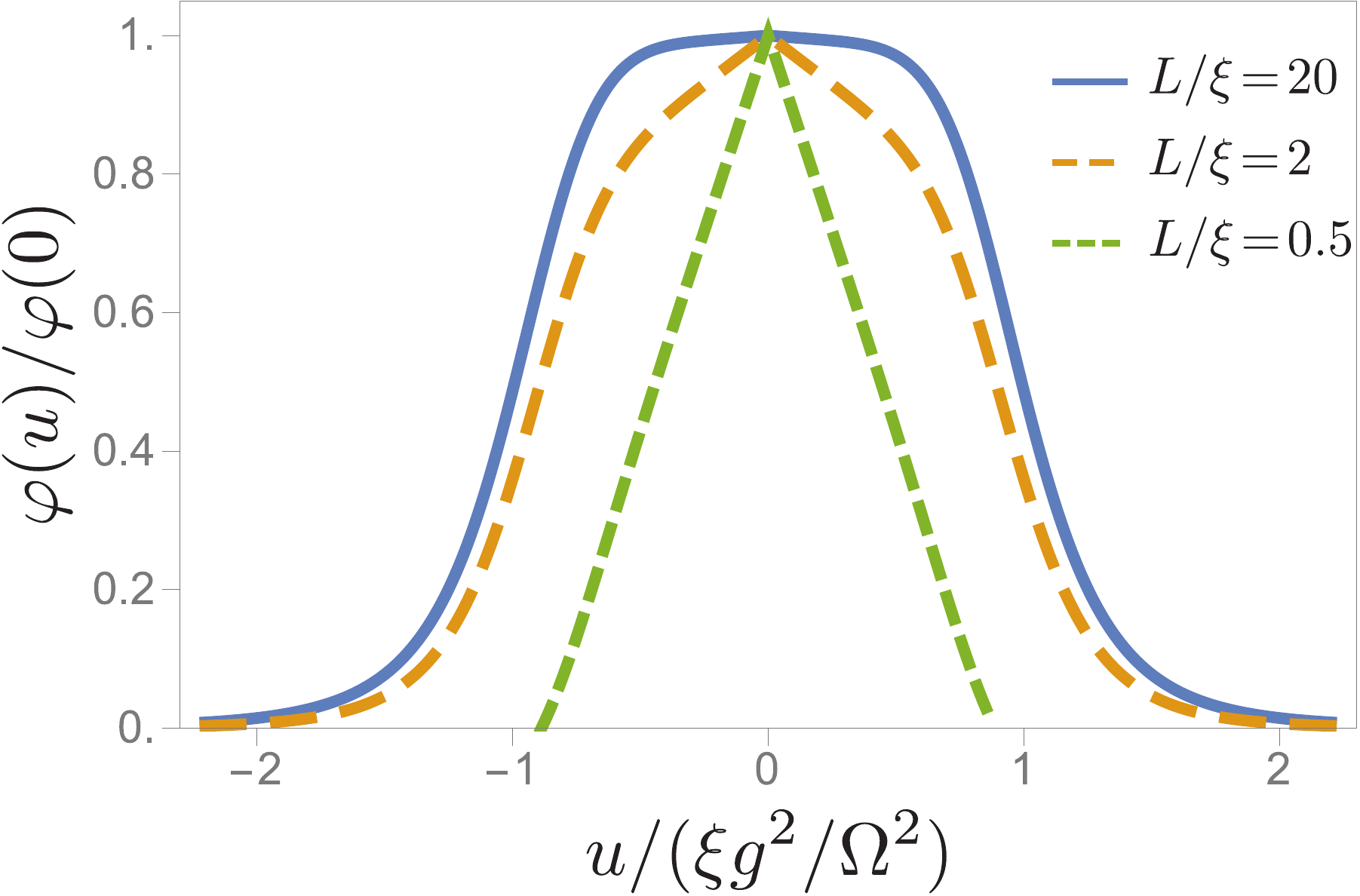}
\includegraphics[width= 0.45\columnwidth]{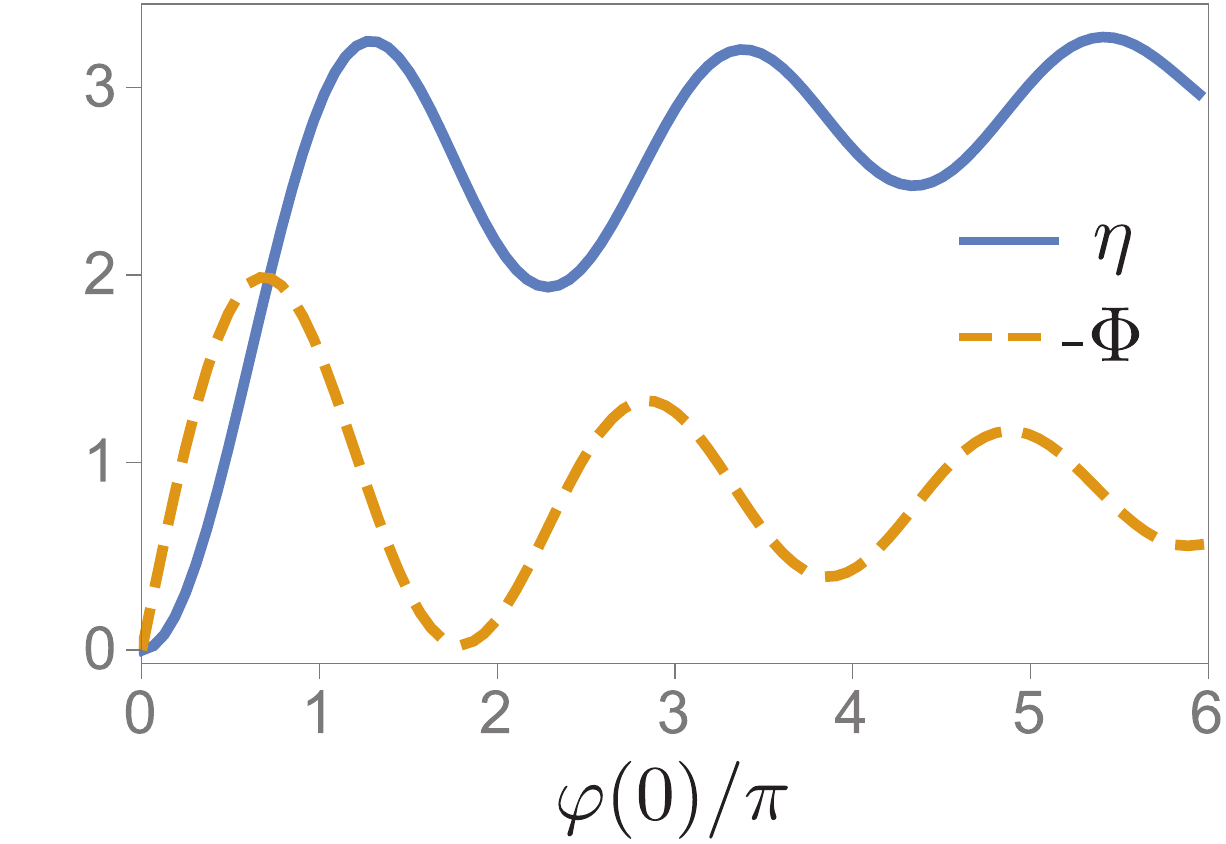}\\
\caption{ %$\xi=14\mu m$ and 
(left) Phase factor $\varphi(u)$ for homogeneous distributions of atoms with different lengths $L$. For short clouds the condition $\xi\ll L$ is not satisfied and the behavior of $\varphi(u)/\varphi(0)$ is no longer universal, like it is for a long medium. (right) Phase shift and suppression of the electric field for coherent state in the limit $\xi_{\rs out} \ll l_{\rs coh}$ as a 
 function of a single photon nonlinearity strength parametrized by $\varphi(0)$.
For weak nonlinearities $\varphi(0)\ll 1$ the electric field suppression scales quadratically and the phase shift linearly with the nonlinearity strength.  
%Maximal suppression takes place when  $\varphi(0) \sim \pi \text{ (mod } 2\pi )$.
The suppression as well as phase shift oscillate with the increasing strength of interaction.
%For $\varphi(0)=\pi/20$ we have $\xi/L=1.58$. 
}
\label{fig2}
\end{figure}

We start by analyzing the behavior of the {\it two-photon} solution. It allows us to
determine the influence of the involved approximations and to provide a connection to
previous results on two-polariton propagation \cite{Firstenberg2013a,Bienias2014b}. For an
arbitrary two photon state $|\phi\rangle$,
the incoming wave function is defined via
\begin{equation}
\phi^{\rs in}(x-c t,y- c t) = \lim_{t \rightarrow - \infty}    \langle 0|  \hat{\psi}(z,t) \hat{\psi}(w,t)   |\phi\rangle/\sqrt{2}, 
\end{equation}
with the coordinates $x=\zeta(z)$ and $y = \zeta(w)$,
and the outgoing wave function $\phi^{\rs out}$ via an analogous expression in the limit $t\ra \infty$. % in the limit $t\rightarrow \infty$,  respectively.
Using the above exact solution for the bosonic field operators $\hat{\psi}(z,t)$, we obtain the relation between 
the incoming and the outgoing photon wave function
\begin{equation}
\phi^{\rs out}(x,y,t) =e^{- i \varphi(x - y)}     \phi^{\rs in}(x-c t' ,y-c t')
\label{twoPhoton}
\end{equation}
where $t' = t-\Delta t$  accounts for the delay of the polaritons inside the medium with $\Delta t = \int_{-\infty}^{\infty} \text{d}y  \left({1}/{ \beta(y)^{2}}-1\right)/c$. Note, that the outgoing
wave function only depends on the reduced coordinate $\tau_1= x -c t'$ and $\tau_2 = y -c t'$; therefore, in the following, we will use these reduced coordinates to express
the outgoing wave function. The phase factor  $ \varphi(u)$
describes the  correlations built up between the photons during the propagation through the medium and takes the form
\begin{equation}
 \varphi(u) =  \frac{1}{\hbar c} \int_{-\infty}^{\infty} \!\!\text{d}w\: \tilde{n}\big(w+u\big) \tilde{ n}\big(w\big)  \tilde{V}\big(w+u,w\big). 
 \label{varphiU}
\end{equation}
It is instructive to analyze this phase factor for a specific homogeneous atomic density distribution 
$ n_{a}(x) = \bar{n}_{a} \theta(L^2/4-x^2)$ where $\theta$ is a Heaviside step function.
The time delay simplifies to  $\Delta t =  L  (1/\vC_{g}-1/c)$, where the slow light velocity $\vC_{g} = c \Omega^2/(g^2+\Omega^2)$
and the collective coupling between photon and matter $g= g_{0} \sqrt{\bar{n}_{a}}$.
In turn, the phase shift acquires the peak value 
\begin{equation}
\varphi(0) = \frac{g^4V(0) L}{(g^2+\Omega^2)\Omega^2 \hbar c} = -\frac{g^2}{g^2+\Omega^2} \frac{ \kappa \gamma}{\Del}
\label{varPhiZero}%\nonumber
\end{equation}
with $\kappa$ the optical depth of the medium. 
The width of the signal in the phase $\varphi(u)$ is enhanced from the blockade radius by the slow light velocity to $ \xi_{\rs out}= \xi (g^2+\Omega^2)/\Omega^2$. The exact phase shapes for different medium lengths are shown in Fig.~\ref{fig2}.
The determination of $\varphi(u)$ for other physical distributions of atoms is straightforward.

Note, that the interaction provides a spatially dependent phase factor correlating the photons, but is unable to induce a
modification in the intensity correlations. A bunching of photons as observed in the experiments by Firstenberg et al.
\cite{Firstenberg2013a} requires the inclusion of the mass term.  
Here, we estimate the influence of this term,
and determine the regime of validity for our approximation to drop it, for details see the supplement material. First, the inclusion of the mass would 
lead to an additional phase shift estimated by 
$\varphi_{\rs m} \sim \hbar \Delta t/ (m \xi^2)|\varphi(0)^2+i\varphi(0)|= |\varphi(0)+i| g^6/(g^2+\Omega^2)^3L^2/\xi^2$, where we used the
expression for the polariton mass $m =\hbar \frac{(g^2+\Omega^2)^3}{ 2 c^2 g^2 \Delta \Omega^2}$ \cite{Bienias2014b}.
In order to drop phenomena like the bunching of photons we require $\varphi_{\rs m}  \ll 1$. Secondly, we would like the 
phase shift induced by the interaction to dominate the behavior, i.e.,  $\ \varphi(0) \gg \varphi_{\rs m}$.
The two conditions are either satisfied for a weak coupling between photon and matter 
or for a short medium.

%{\bf $N$-photon state:} 
The two-photon analysis can be generalized straightforwardly to an $N$-photon Fock state.
Then, the wave function reduces to
\begin{eqnarray}
   \nn   \frac{\phi^{\rs out}(\tau_{1},\ldots, \tau_{N})}{\phi^{\rs in}(\tau_1,\ldots,\tau_{N})} = \exp\left[- i \sum^N_{i < j}  \varphi(\tau_{i}- \tau_{j})\right]\nonumber
   \label{eq:Nphoton}\nn,
\end{eqnarray}
%
%
%\begin{eqnarray}
 %  \nn   \frac{\phi^{\rs out}(x_{1},\ldots, x_{N},t)}{\phi^{\rs in}(x_1\!-\! c t',\ldots,x_{N}\!-\!c t')} = \exp\left[- i \sum^N_{i < j}  \varphi(x_{i}- x_{j})\right]\nonumber
 %  \label{eq:Nphoton}\nn,
%\end{eqnarray}
%
where $\tau_i = x_i- c [t-\Delta t]$.
This allows one to derive the outgoing wave function for an arbitrary incoming state. Of special experimental interest  however is the behavior of
coherent states.  A general incoming  coherent state is characterized by its incoming
electric field expectation value $\alphaIn(x-ct) = \lim_{t\rightarrow-\infty}E(x,t)$ with $E(x,t)=  \langle\alphaIn |  \psi(x,t) \beta(x) |\alphaIn\rangle $.
Then, the outgoing electric field behaves as 
\begin{equation}
\frac{E^{\rs out}(\tau)}{  \alphaIn(\tau)}= \exp\left(\int \text{d}u |\alphaIn(u)|^2 \left[e^{-i \varphi(u - \tau)} - 1\right] \right),  \label{coherentstate}
\end{equation}
where $\tau = x- c [t-\Delta t]$.
%
%
%\begin{eqnarray}
%\frac{E^{\rs out}(x,t)}{\alphaIn(x\!-\! c t')}=   \exp\left(\int du |\alphaIn(u)|^2 \left[e^{-i \varphi(u - x+c t')} - 1\right] \right),\nn
%\end{eqnarray}
%
In the limit of a weak non-linearity  $\varphi(u) \ll 1$, we can recover the result of a classical Kerr nonlinearity. In this regime, the incoming wave packet has a size $l_{\rs coh}$ much larger than the characteristic size of the
interaction $l_{\rs coh}\gg \xi_{\rs out}$, and propagates through a long medium $L\gg \xi$. Then, the Eq.~(\ref{coherentstate}) reduces to
$
E^{\rs out}(\tau)= \alphaIn(\tau)  \exp\left(- i \: \sigma \: |\alphaIn(\tau)|^2\right)
$ %\end{equation}
with  $\sigma$ the strength of the Kerr nonlinearity. The latter depends on the shape of
the atomic density distribution and reduces for a homogeneous atomic density to 
\begin{equation}
% \sigma = \int du \:  \varphi(u) = \frac{2\pi}{3\sqrt{2}}  \frac{  \kappa \gamma \xi}{ \Del} \frac{g^2}{ \Omega^2}  .
 \sigma = \int \text{d}u \:  \varphi(u) = \frac{2\pi}{3} \frac{g^2}{\Omega^2+g^2} \frac{  \kappa \gamma}{ \Del}\xi_{\rs out}.
\end{equation}
However,  it is important to stress, that  Eq.~(\ref{coherentstate}) includes also the corrections to the Kerr nonlinearity  due to the quantum fluctuations. %, and therefore, 
%provides also the corrections to a classical 
% by quantum fluctuations. 
The corrections can be analyzed
by the full evaluation of the factor $i\Phi +\eta= - \integral{u}(\exp(-i\varphi(u))-1)/ \xi_{\rs out}$, where $\Phi$ describes the strength
of the Kerr nonlinearity, whereas $\eta$ accounts for a suppression of the coherences due to quantum fluctuations, see Fig.~\ref{fig2}.
The latter follows from the fact that a coherent state is a superposition of different number states, where each number state picks 
up a slightly different phase factor.

The full characterization of the output state and relation to experimentally accessible quantities is most conveniently 
achieved by the  normally ordered electric field correlations in the reduced coordinates $\tau_i$, 
% $\tau_i=x_{i}-c [t -\Delta t]$,
%
\begin{eqnarray}
G^{\rs out}_{n,m}( \tau_1,...,\tau_{n+m})=%\qquad\qquad\nn\\
\left\langle
\alphaIn\left|
{\prod_{i= 1}^n\psi\dr(\tau_i) \prod_{j=n+1}^{n+m}\psi(\tau_j)}
\right|{\alphaIn}
\right\rangle. \nn
 \label{gnm}\label{Gnm}
 \end{eqnarray}
 These correlation functions are experimentally accessible in a homodyne detection 
 scheme. The full expression for the correlations of the outgoing fields for an incoming coherent state 
 is presented in the supplement. In the following, we provide the result for the two point 
 correlation function $G^{\rs out}_{0,2}$, which reduces to
 \begin{eqnarray}
 G^{\rs out}_{0,2}(\tau,\tau')
& = &  \alphaIn(\tau) \alphaIn(\tau') \exp\left[- i \varphi(\tau-\tau')\right]
\label{G02}\\
&&\hspace{-25pt}  \times \exp\left(\integral{u} |\alphaIn(u)|^2 \left[e^{- i \varphi(u -\tau )- i \varphi(u-\tau')}-1\right] \right). \nonumber
\end{eqnarray}
We can distinguish two different contribution: first, we find a strong spatial correlation determined by the phase contribution 
$\varphi(\tau-\tau')$, which provides direct information about the effective interaction potential between the polaritons. It is this
contribution, which allows the access to  the effective interaction potential within a homodyne detection scheme. 
The last factor
describes additional phase shift and the suppression %\todoc{it is not just suppression, maybe that it is weaker for $l_{\rs coh}\gg \xi$} 
due to quantum fluctuations, which are small corrections for $\xi_{\rs out}  |\alphaIn(\tau)|^2 \ll 1$.

A full characterization of the outgoing field for an incoming field coherent field $\alphaIn$ is provided by the Wigner function
 $W(q,p)$. In contrast to circuit and cavity QED experiments, where the
photons within the resonator are characterized by a single photonic mode \cite{Wallraff2004,Haroche2006}, our system here corresponds to a multimode setup. 
Therefore,  in terms of the Wigner function, we can only express the reduced density matrix in a specific photonic mode. 
For this purpose, we define the annihilation operator for an arbitrary spatial mode $u(x)$ as
$\aOp = \integral{x} u(x) {\psi}(x)$
and the related quadrature operators as $\hat{q}=(\aOp+\aOp^\dagger)/2$, $\hat{p}=(\aOp-\aOp^\dagger)/2i$.
Then, the Wigner function derives directly from the analytical expression for the correlation functions $G^{\rs out}_{n, m}$ for the incoming coherent field~\cite{suppBienias},
\begin{equation}
W(q,p)=\frac{2}{\pi}\sum_{nm}\frac{(-1)^{n+m}}{n!m!}\corr_{nm}\partial_{\alpha*}^n\partial_{\alpha}^me^{-2|\a|^2}
\end{equation}
with $\a=q+i p$,  and $\corr_{nm}$ the overlap of the electric field correlations with the probe photonic mode
\begin{eqnarray}
\corr_{nm}\!=\!\integral{^{n+m}\tau}G^{\rs out}_{n,m}(\tau_1,...,\tau_{n+m})\prod_{i=1}^{n} u(\tau_i)^{*}\!\! \prod_{j=n+1}^{m+n} u(\tau_j)\nn.
\end{eqnarray}
\begin{figure}%[H]
\includegraphics[width= 0.49\columnwidth]{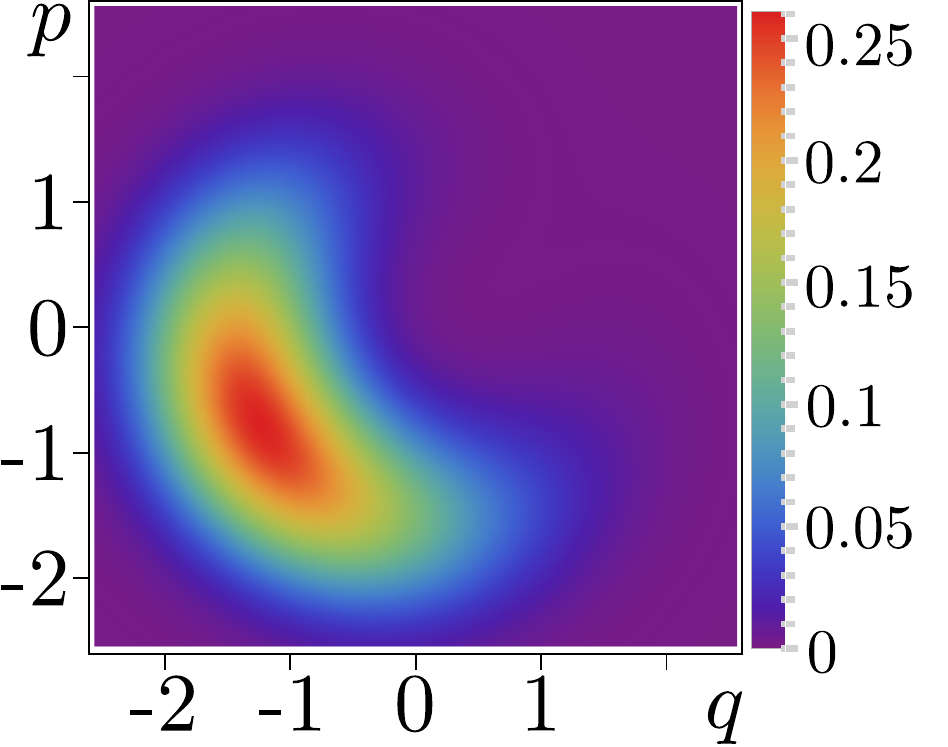}
\includegraphics[width= 0.49\columnwidth]{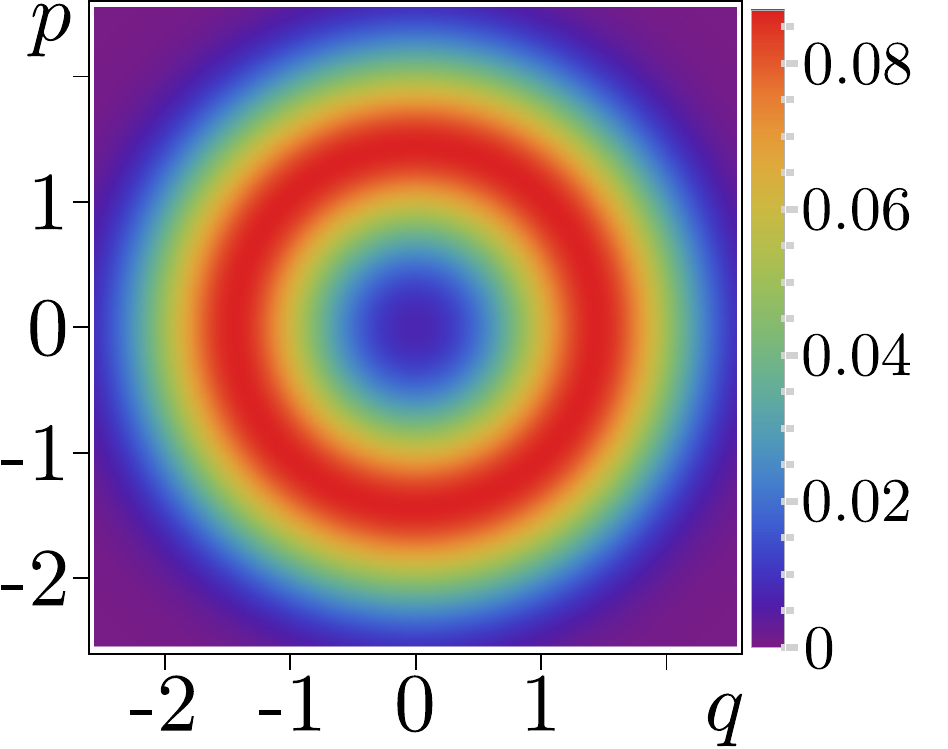}
\caption{ 
 Wigner function describing short range correlations ($l_{\rs probe}\ll\xi_{\rs out}$) for long-photons
 $l_{\rs coh}/l_{\rs probe}=100$
%for the Gaussian distribution of atoms 
for two different strengths of interaction: $\varphi(0)=\pi/64$ (left) and $\varphi(0)=\pi$ (right).
}
\label{fig3}
\end{figure}
In order to characterize short range correlations between photons we consider a homodyne detection~\cite{Deleglise2008,Lvovsky2009a,Bimbard2014} with $u(x)$ being a localized mode having  size $l_{\rs probe}$ much shorter than  $\xi_{\rs out}$.
The quasi-probability $W(q,p)$ for different strengths of the interaction is shown in Fig.~\ref{fig3}.
For weak interactions $\varphi(0)\ll 1$, the leading correction due to quantum fluctuations to the Gaussian coherent state is
a small squeezing. However, for increasing interaction we obtain a strongly mixed state. % due to the strong spatial entanglement caused by the interaction. 
This behavior is a result of the localized measurement tracing out all positions outside the $u(x)$. 
Such an operation, acting on our strongly spatially entangled state, leads to the mixed state.
%For strong interactions $\varphi(0)\,\sigma/\sigma_f\sim \pi$ the large effect is visible in the form of the undefined phase of the output state,  Fig.~\ref{fig10} (b).

A crucial property of our analysis is that it demonstrates the possibility to probe the microscopic interaction potential between the Rydberg polaritons via
a homodyne detection scheme for a coherent input state. This method can easily be extended to probe higher body interactions between the polaritons, which are
expected to appear for higher polariton densities. Such a $n$-body interaction on the microscopic level takes the from
\begin{equation}
H_{n}= \frac{1}{n!} \int \!\! \text{d}{\bf x} \: U_{n}(x_{1}, \ldots, x_{n})\: \prod_{i=1}^{n}  n(x_i)  \psi^{\dag}(x_i) \psi(x_i)
\end{equation}
with the $n$-body interaction potential $U_{n}$. This term can be straightforwardly included in the exact solution. As an example, we present the results for a three-body interaction,
which leads, in analogy to Eq.~(\ref{twoPhoton}), to  a phase contribution to 
the three photon wave-function 
\begin{eqnarray}
   \phi^{\rs out}(\tau_1,\tau_2,\tau_3)=e^{- i \varphi_3(\tau_1-\tau_2,\tau_2-\tau_3)}     \phi^{\rs in}(\tau_1 ,\tau_2,\tau_3 )\nn.
\end{eqnarray}
The phase factor $\varphi_3(u,v)$ induced by the three-body interaction 
takes the form
\begin{eqnarray}
%=\hspace{19em}\nn \\
 %\hspace{1em}  
 \frac{1}{\hbar c} \int_{-\infty}^{\infty} \!\!\text{d}w\: \tilde{n}\big(w\!+\!u\big) \tilde{n}\big(w\!+\!v\big) 
 \tilde{ n}\big(w\big)  \tilde{U}_{3}\big(w\!+\!u,w\!+\!v,w\big) \nonumber,
\end{eqnarray}
with $ \tilde{U}_{3}$ defined in analogy to $\tilde{V}$. The corresponding three-body interaction potential can then be experimentally observed
in a homodyne detection of the correlations  $G^{\rs out}_{0,3}$.

In conclusion,  we studied Rydberg slow light polaritons and obtained a consistent quantum theory for a Kerr nonlinearity. 
For weak interactions we demonstrated that the system reduces to a conventional Kerr nonlinearity, while
for moderate interactions we derived quantum corrections. 
Rydberg slow light polaritons naturally lead to a finite interaction range and a mass term, which regularize previous problems in deriving a quantum theory of
a Kerr nonlinearity. %Here, the difference is that
Our approach provides a promising tool for the direct experimental observation of the two-body interaction potential as well as higher body interaction potentials via a homodyne detection.

We thank S. Hofferberth and A.V. Gorshkov  for discussions. We acknowledge the European Union H2020 FET 
Proactive project RySQ (grant N. 640378), and support by the Deutsche Forschungsgemeinschaft (DFG) 
within the SFB/TRR 21.

%%%%%%%%%%%%%%%%%%%%%%%%%%%%%%%%%
%\bibliographystyle{/Users/admin/Dropbox/notes/latex/bernd}
%\bibliography{/Users/admin/Documents/Bibtex/library}

\bibliographystyle{bernd}
\bibliography{library}

\clearpage
\newpage
\begin{widetext}
\section{Supplemental material}

% !TeX root =Quantum-Kerr-effect_hpb2016.tex
%=================%
%
%=================%

\beginsupplement
\setcounter{equation}{0}
\renewcommand{\theequation}{S\arabic{equation}}
%{ \large \centering Supplemental material for:  ``Quantum theory of Kerr nonlinearity with Rydberg slow light polaritons''}

\subsection{Regime of parameters in which mass term is negligible}%: derivation
In this section, we derive a regime of parameters in which we can drop the mass term in the polaritonic dispersion relation.
For this purpose, we analyze two polaritons propagating through a cloud of atoms with a constant density. % and length $L$.
In the relative $r$ and center of mass  $R$ coordinates, the  \Schroedinger equation  for the two-body wave function $\psiTwoB(R,r)$ takes the form \cite{Bienias2014b} 
\begin{eqnarray}
\hbar\omega\psiTwoB(R,r)= \left(-i\hbar v_g\partial_R-\frac{\hbar^2}{m}\partial_r^2+\alpha V(r)\right)\psiTwoB(R,r),
\label{eq1aa}
\end{eqnarray}
where
\begin{eqnarray}
m =\hbar \frac{(g^2+\Omega^2)^3}{ 2 c^2 g^2 \Delta \Omega^2},\hspace{5em} 
V(r)=\frac{2\Omega^2\hbar}{\delta}\frac{1}{1+r^6/\xi^6}, \hspace{5em} 
\alpha = \frac{g^4}{(g^2+\Omega^2)^2}, \hspace{5em} 
 v_g=\frac{\Omega^2}{\Omega^2+g^2}c.
\end{eqnarray}

Assuming that the mass term is negligible, we can find the analytical solution of  Eq.~\ref{eq1aa} of the form
\begin{eqnarray}
\psiTwoB_0(R,r)%=\psi(0,r)e^{i\omega/v_g R}\exp(-i\alpha RV(r)/v_g) 
=\psiTwoB(0,r)
\exp
\left(i\frac{\omega}{v_g} R-i\phiTwo(0)\frac{R}{L}  \frac{1}{1+(r/\xi)^6}
\right) 
%=\psi(0,r)\exp\left(-i{\phi_{int}(R,r)}\right)
\end{eqnarray}
where we used the relation
$\phiTwo(0)=L\alpha  V(0)/v_g$, see Eq. \suppref{varPhiZero}9. For latter purposes let us define $\phiTwo_{int}(R,r)=\phiTwo(0)R/L/(1+(r/\xi)^6)$. % and index ``$_{0}$'' in $\Psi_0$ denotes the zeroth order in the mass term. % whereas position dependent phase factor  $\phi_{int}(R,r)=\phi(0)\frac{R}{L}  \frac{1}{1+(r/\xi)^6}$.
Using the above solution, we calculate perturbatively
the corrections due to the mass term to the phase shift.
 For this purpose we express the full solution $\psiTwoB$ using $\psiTwoB_0$: % and  the exponent $\phiTwoMass_m(R,r)$:
 \begin{eqnarray}
\psiTwoB(R,r)=\psiTwoB_0(R,r) e^{-i\phiTwoMass_m(R,r)},
\end{eqnarray}
where $\phiTwoMass_m(R,r)$ takes into account the impact of the mass term.
Next, we insert this Ansatz  into Eq.~\ref{eq1aa}. Exploiting that  $\psiTwoB_0$ is the solution for the unperturbed Hamiltonian, most of the terms cancel and we arrive with the equation for $\phiTwoMass_m$:
\begin{eqnarray}
0= -iv_g\psiTwoB_0(R,r)\partial_Re^{-i\phiTwoMass_m(R,r) }-\frac{\hbar^2}{m}\partial_r^2\psiTwoB_0(R,r) e^{-i\phiTwoMass_m(R,r) }.
\label{eqForPhiM}
\end{eqnarray}  
This equation can be simplified once we 
take into account that in the perturbative  limit  $|\partial_r\phiTwoMass_m(R,r)| \ll |\partial_r\phiTwo_{\rs int}(R,r)| $ and $|\partial_r^2\phiTwoMass_m(R,r)| \ll |\partial_r^2\phiTwo_{\rs int}(R,r)|$. Moreover, considered photons are much longer than $\xi_{\rs out}$ and, therefore, we drop $\partial_r\psiTwoB(0,r)$ and $\partial_r^2\psiTwoB(0,r)$ terms. 
Finally,  Eq. \ref{eqForPhiM} simplifies to
\begin{eqnarray}
0= -iv_g\psiTwoB_0(R,r)\partial_Re^{-i\phiTwoMass_m(R,r) }-\frac{\hbar^2}{m}\psiTwoB(0,r)e^{-i\phiTwoMass_m(R,r) }
e^{i\frac{\omega}{v_g} R}\partial_r^2
\exp
\left(-i\phiTwo(0)\frac{R}{L}  \frac{1}{1+(r/\xi)^6}
\right).
\label{}
\end{eqnarray}  
This equation leads to the solution for $\phiTwoMass_m(L,r)$  of the form
%e^{-i\phi_m }
\begin{eqnarray}
\phiTwoMass_m(L,r)
=-
\frac{1}{v_g} \integralb{0}{L}{R}\frac{\hbar^2}{m}
\frac{\partial_r^2\exp\left(-i\phiTwo(0)\frac{R}{L} \frac{1}{1+(r/\xi)^6}
\right) }
{\exp\left(-i\phiTwo(0)\frac{R}{L} \frac{1}{1+(r/\xi)^6}
\right)  }.
%=-\frac{1}{v_g} \integralb{0}{L}{R}\frac{\hbar^2}{m}
%\frac{\partial_r^2\exp(-i\alpha R V_e(r)/v_g)}{\exp(-i\alpha R V_e(r)/v_g) }
\end{eqnarray}
%
%Expression for $\phi_m(R=L,r)$ is $r$ dependent, and 
In order to estimate $\phiTwoMass_m(R=L,r)$ we consider its value at $r=\xi$, which is equal to
\begin{eqnarray}
\phiTwoMass_m(L,\xi)=3  \left(\text{$\phiTwo $(0)}+i\right)\frac{  L^2 }{\xi ^2 }\frac{g^6}{\left(g^2+\Omega ^2\right)^3},
\end{eqnarray}
and corresponds to the result  for $\varphi_m$ from the main text.
%\todoc{Convention with $\phi,\varphi$}
%
%\subsection{Closer look at experimental conditions}
We see that the mass term can be dropped if two conditions $\varphi_m\ll 1$ and $\varphi_m/\varphi(0)\ll 1$ are satisfied.
%Next, we we show the implications of these two conditions in two regimes: $\xi\gg L$ and $\O\gg g$.

%In the case of $\xi\gg L$ mass term can be neglected for intermediate interaction strengths such that  $L^2/\xi^2 \text{max}(|\varphi(0)|,1/|\varphi(0)|)\ll 1$ is satisfied. Note, that in this regime always $g\gg\O$.

\subsection{Correlations of the outgoing fields for an incoming coherent state}
Here, we derive the general expression for the correlations $G^{\rs out}_{n,m}$ of the outgoing fields for an incoming coherent state $\ket{\alphaIn}$. 
We start by %expressing the interacting field operators $\hat{\psi}$ using the noninteracting field operators $\hat{\psi}_0$  [i.e., 
inserting  the exact solution for bosonic field operators 
$ 
\hat{\psi}(z,t) = e^{-i  \hat{J}(z,t)  } \hat{\psi}_{0}(z-c t)
$
into the definition of $G^{\rs out}_{n,m}$ from the main text. This leads to
\begin{eqnarray}
G^{\rs out}_{n,m}( \tau_1,...,\tau_{n+m})=%\qquad\qquad\nn\\
\left\langle
\alphaIn\left|
{\prod_{i= 1}^n%\psi\dr(\tau_i) 
\left(e^{-i  \hat{J}(z_i,t)  } \hat{\psi}_{0}(z_i-c t) \right)\dr
\prod_{j=n+1}^{n+m}
e^{-i  \hat{J}(z_j,t)  } \hat{\psi}_{0}(z_j-c t) 
}
\right|{\alphaIn}
\right\rangle 
 \label{GnmAA},
 \end{eqnarray}
where $\tau_i=z_i-ct$.
Our goal is to transform the product of operators to the normally ordered expression, of which expectation value in a coherent state is trivial. 
For this purpose, we first use the relation 
\beqa
\hat{\psi}_{0}(z_i-c t)e^{-i  \hat{J}(z_j,t)  }=
e^{-i  \hat{J}(z_j,t)  } e^{- i \varphi(z_i - z_j)} 
 \hat{\psi}_{0}(z_i-c t)
\eeqa
%which can be proofed using BCH relation
% e^{X}Y e^{-X}  =Y+\left[X,Y\right]+\frac{1}{2!}[X,[X,Y]]+\frac{1}{3!}[X,[X,[X,Y]]]+\cdots.
% to move operators $\Psi_0$ to the sides of the full expression/ \\
to normally order the $\psi_0$ operators in the Eq.~\ref{GnmAA},
\begin{eqnarray}
G^{\rs out}_{n,m}( \tau_1,...,\tau_{n+m})&=&%\qquad\qquad\nn\\
\left\langle
\alphaIn\left|
\prod_{k= 1}^n\hat{\psi}\dr_{0}(\tau_k) 
\prod_{i= 1}^n%\psi\dr(\tau_i) 
e^{i  \hat{J}(z_i,t)  }
\prod_{j=n+1}^{n+m}
e^{-i  \hat{J}(z_j,t)  } 
\prod_{l=n+1}^{n+m}
\hat{\psi}_{0}(\tau_l) 
\right|{\alphaIn}
\right\rangle \nn\\
&\times&\exp\left[i \sum_{k>l=1}^n\varphi(\tau_k-\tau_l)\right]
\exp\left[-i \sum_{k>l=n+1}^{m+n}\varphi(\tau_k-\tau_l)\right].
\label{GnmXX}
 \end{eqnarray} 
%
%Since,  
Next, we use the fact that in the limit of $t\ra \infty$ the expression for $\hat{J}(z_i,t)$ % [defined in Eq. \ref{Jzt}] 
can be written as
%\beqa
$
\hat{J}(z_i,t) = \int^{-\infty}_{\infty}\text{d}u\: \hat{ I}(u)  \varphi(u-z_i+ct),
$
%\eeqa
%
and that $\hat{J}(z_j,t)$ commutes with $\hat{J}(z_i,t)$, in order to rewrite the product of exponentials 
%$\exp(\pm i\hat{J}(z_i,t))$ terms 
appearing in Eq. \ref{GnmXX} in the following way:
\begin{eqnarray}
\prod_{i= 1}^n%\psi\dr(\tau_i) 
e^{i  \hat{J}(z_i,t)  }
\prod_{j=n+1}^{n+m}
e^{-i  \hat{J}(z_j,t)  } 
=
\exp
\left(i\sum_{i= 1}^n%\psi\dr(\tau_i) 
  \hat{J}(z_i,t)  
-i\sum_{j=n+1}^{n+m}
  \hat{J}(z_j,t)  \right)
\nn
\\
=
\exp\left(\integralb{-\infty}{\infty}{u} \hat{I}(u) \left[
i\sum_{i= 1}^n \varphi(u -\tau_i )-i\sum_{j=n+1}^{n+m}\varphi(u-\tau_j)
\right] \right).
    \end{eqnarray}
The last expression can be transformed to the normally ordered one using the relation~\cite{Boivin1994}: %, see Ref.
\begin{eqnarray}
\exp\left(
\integral{u} g(u) \hat{I}(u)
\right)
=\normord{
\exp
\left(\integral{u}
\left(e^{g(u)}-1\right)\hat{I}(u)\right)
}.
\end{eqnarray}
%which can be found in Ref. \cite{Boivin1994}, 
In our case $g(u)=i\sum_{i= 1}^n \varphi(u -\tau_i )-i\sum_{j=n+1}^{n+m}\varphi(u-\tau_j)$, what leads to
%as well as that $\hat{J}(z_j,t)$     and $\hat{J}(z_i,t)$ commute, we transform the product of $\exp(\pm i\hat{J}(z_i,t))$ terms appearing in Eq. \ref{GnmXX} into the normally ordered expression:
\begin{eqnarray}
\prod_{i= 1}^n%\psi\dr(\tau_i) 
e^{i  \hat{J}(z_i,t)  }
\prod_{j=n+1}^{n+m}
e^{-i  \hat{J}(z_j,t)  } 
=
:\exp\left(\integralb{-\infty}{\infty}{u} \hat{I}(u) \left[
\exp\left(
i\sum_{i= 1}^n \varphi(u -\tau_i )-i\sum_{j=n+1}^{n+m}\varphi(u-\tau_j)
\right)-1\right] \right)
:.
    \end{eqnarray}
The last equation inserted into the Eq. \ref{GnmXX} provides  the final result,
\begin{eqnarray}
G^{\rs out}_{n,m}( \tau_1,...,\tau_{n+m})&=&
\prod_{i= 1}^n\hat{\alphaIn}^*(\tau_i) \prod_{j= n+1}^{n+m}\hat{\alphaIn}(\tau_j) 
\exp\left[i \sum_{k>l=1}^n\varphi(\tau_k-\tau_l)\right]
\exp\left[-i \sum_{k>l=n+1}^{m+n}\varphi(\tau_k-\tau_l)\right]\nn\\
&\times&\exp\left(\integralb{-\infty}{\infty}{u} |\alphaIn(u)|^2 \left[
\exp\left(
i\sum_{i= 1}^n \varphi(u -\tau_i )-i\sum_{j=n+1}^{n+m}\varphi(u-\tau_j)
\right)-1\right] \right).
 \end{eqnarray} 
Two special cases of these correlations, i.e., $G^{\rs out}_{0,1}$ and $G^{\rs out}_{0,2}$ are presented in the main text, see Eqs \suppref{coherentstate}10 and \suppref{G02}12, respectively.

%%%%%%%%%%%%%%%%%%%%%%%%%%%%%%%%

\subsection{Wigner function from correlation functions}
Here, we show how Wigner function  $W(q,p)$ can be calculate using the correlation functions $\corr_{nm}$.
Our starting point is symmetrically ordered characteristic function $\chi(\eta)$ defined as
\begin{equation}
\chi(\eta) = \text{Tr} \left[ \rho \exp\left[\eta \aOp\dr-\eta^*\aOp \right]\right].
\end{equation}
The function $\chi(n)$ can be expressed using correlation function $\corr_{nm}=\meanv{(\aOp\dr)^n\aOp^m}$ as 
\begin{eqnarray}
\chi(\eta) = \sum_{nm}\frac{\eta^n(-\eta^*)^m}{n!m!}e^{-|\eta|^2/2}\text{Tr}\left[\rho (\aOp\dr)^n\aOp^m\right]=\sum_{nm}\frac{\eta^n(-\eta^*)^m}{n!m!}
e^{-|\eta|^2/2} \corr_{nm}.
\label{chiEta}
\end{eqnarray}
The Wigner function is defined as  the Fourier transform of the
 characteristic function $\chi(\eta)$~\cite{Walls2008},
\begin{eqnarray}
W(\a) &=& \frac{1}{\pi^2}\integral{^2\eta} e^{\eta^*\a-\eta\a^*}\chi(\eta).
\label{WDef}
\end{eqnarray}
Finally, we insert $\chi(\eta)$ from Eq. \ref{chiEta} into the definition \ref{WDef} and afterwards transform $W(\alpha)$ to a more concise expression:
\begin{eqnarray}
W(\alpha)&=& \frac{1}{\pi^2}\sum_{nm}\frac{(-1)^m}{n!m!}\integral{^2\eta} e^{\eta^*\a-\eta\a^*}\eta^n(\eta^*)^m e^{-|\eta|^2/2}\corr_{nm}\nn\\
&=& \frac{1}{\pi^2}\sum_{nm}\frac{(-1)^{n+m}}{n!m!}\corr_{nm}
\partial_{\a^*}^n\partial_{\a}^m
\integral{^2\eta} e^{\eta^*\a-\eta\a^*}e^{-|\eta|^2/2}\nn\\
&=& 
\frac{2}{\pi}\sum_{nm}\frac{(-1)^{n+m}}{n!m!}\corr_{nm}
\partial_{\a^*}^n\partial_{\a}^m
e^{-2|\a|^2},
\end{eqnarray}
which is the formula presented in the main text.

\newpage 
\end{widetext}

\end{document}